\documentclass[twoside,fleqn]{ActaStyle}
\usepackage{times}
\usepackage{epsfig}

\def\beq{\begin{equation}}
\def\eeq{\end{equation}}

\def\bea{\arraycolsep .1em \begin{eqnarray}}
\def\eea{\end{eqnarray}}
\def\Tr{{\rm Tr}}

\def\de{\delta}

\def\eq#1{(\ref{#1})}

\def\Eq#1{Eq.~(\ref{#1})}

\def\s0#1#2{\mbox{\small{$ \frac{#1}{#2} $}}}
\def\0#1#2{\frac{#1}{#2}}

\def\grgl{\:\hbox to -0.2pt{\lower2.5pt\hbox{$\sim$}\hss}{\raise3pt\hbox{$>$}}\:}
\def\klgl{\:\hbox to -0.2pt{\lower2.5pt\hbox{$\sim$}\hss}{\raise3pt\hbox{$<$}}\:}

\def\authorheading{Daniel Litim}

\def\shorttitle{Convergence and Stability of the Renormalisation Group}

\begin{document}
\pagerange{1}{8}

\title{CONVERGENCE AND STABILITY OF THE RENORMALISATION GROUP \footnote{Invited talk presented at RG2002, March 10-16, 2002, Strba, Slovakia. E-mail address: Daniel.Litim@cern.ch}}

\author{Daniel F. Litim}
{Theory Division, CERN, CH -- 1211 Geneva 23.}

\vspace*{-5.7cm}
\begin{flushright}
{\normalsize CERN-TH/2002-197}
\end{flushright}
\vspace*{4.5cm}

\abstract{Within the exact renormalisation group approach, it is shown
  that stability properties of the flow are controlled by the choice
  for the regulator. Equally, the convergence of the flow is enhanced
  for specific optimised choices for the regularisation. As an
  illustration, we exemplify our reasoning for $3d$ scalar theories at
  criticality. Implications for other theories are discussed.}

\pacs{11.10.Hi}

\section{Introduction}\label{Introduction}

Renormalisation group techniques are important tools to describe how
classical physics is modified by quantum fluctuations. Integrating-out
all quantum fluctuations provides the link between the classical
theory and the full quantum effective theory
\cite{Zinn-Justin:1989mi}.  A useful method is given by the Exact
Renormalisation Group (ERG) \cite{Bagnuls:2000ae}, which is based on
the Wilsonian idea of integrating-out infinitesimal momentum shells.
ERG flows have a simple one-loop structure. They admit
non-perturbative truncations and are not bound to weak coupling.

An application of the ERG requires some approximations like the
derivative expansion or expansions in powers of the fields. It has
been known since long that approximations induce a spurious dependence
on the regularisation
\cite{Ball:1995ji,Litim:1997nw,Freire:2001sx,Litim:2001hk,Latorre:2000qc}.
This is somewhat similar to the scheme dependence within perturbative
QCD, or within truncated solutions of Schwinger-Dyson equations. While
this scheme dependence should vanish at sufficiently high order in the
expansion, practical applications are always bound to a finite order,
and hence to a non-vanishing scheme dependence.  A partial
understanding of the interplay of approximations and scheme dependence
has been achieved previously. For scalar QED \cite{ScalarQED}, the
scheme dependence in the region of first order phase transition has
been studied in \cite{Litim:1997nw,Freire:2001sx}. For $3d$ scalar
theories, the interplay between the smoothness of the regulator and
the resulting critical exponents has been addressed in
\cite{Liao:2000sh} using a minimum sensitivity condition.  The weak
scheme dependence found in these cases suggests that higher order
corrections remain small, thereby strengthening the results existing
so far.

In this contribution, we review how the convergence and stability of
ERG flows is optimised, thereby providing improved results already to
low orders within a given approximation
\cite{Litim:2000ci,Litim:2001up,Litim:2001fd,Litim:2001dt,Litim:2002cf}.
This involves a discussion on the origin of the spurious scheme
dependence, and its link with convergence and stability properties of
truncated ERG flows. We exemplify the basic reasoning for the
universality class of $O(N)$ symmetric scalar theories in three
dimensions.  It is expected that insights gained from this
investigation will also prove useful for applications to more complex
scalar theories, gauge theories \cite{Litim:1998nf} or gravity
\cite{QuantumGravity1}, which are more difficult to handle.

\section{Renormalisation group flows and truncations}\label{Truncations}

\begin{figure}[t]
\begin{center}
\vskip-.5cm
\unitlength0.001\hsize
\begin{picture}(1000,450)
\put(0,420){ \bf a) full flow}
\put(570,420){ \bf b) truncated  flow}
\put(180,370){ $\bf \Large \Gamma_{\Lambda}$}
\put(750,370){ $\Large \bf \Gamma_{\Lambda}$}
\put(190,-35){ $\Large \bf \Gamma$}
\linethickness{2mm}
\put(740,-35){ $\Large \bf \Gamma^{\rm trunc}$}
\put(280,250){ $ \huge \bf \partial_t\Gamma$ }
\put(850,250){ $ \huge \bf \partial_t\Gamma^{\rm trunc}$}
\put(140,170){ $ \Large \bf R_k$ }
\put(700,170){ $ \Large \bf R_k$ }
\put(460,340){ $ \Large \bf k=\Lambda$}
\put(460,10){ $\Large \bf k=0$}
\psfig{file=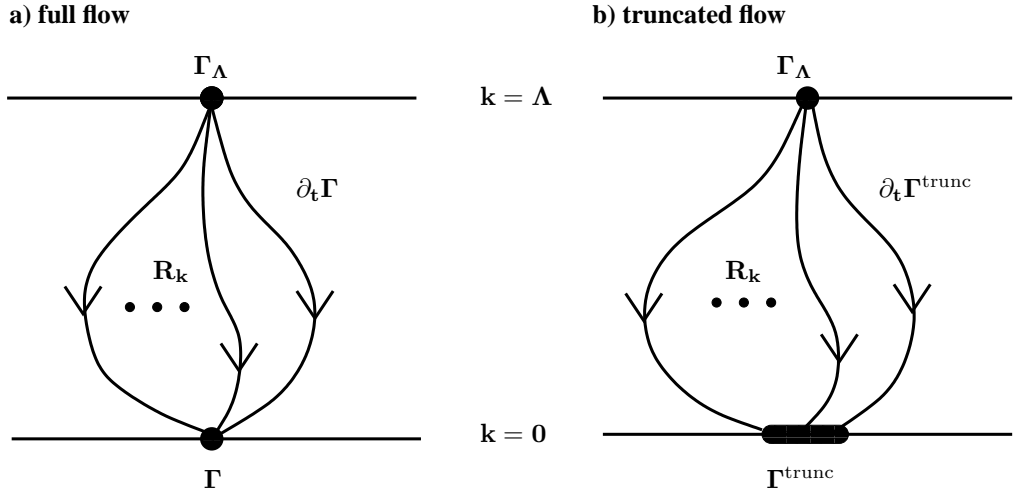,width=\hsize}
\end{picture}
\vskip.3cm
\caption{\label{fig:flows} 
  Schematic diagram of a full (left panel) or truncated (right panel)
  renormalisation group flow connecting an initial effective action at
  $k=\Lambda$ with the full (truncated) quantum effective action at
  $k=0$. The upper line corresponds to the space of all initial
  effective actions. The lower line corresponds to the space of
  effective actions. For $k\neq 0$, the flow trajectories depend on
  the regulator $R_k$. For the full flow, all trajectories join at
  $k=0$. For a truncated flow, the endpoint depends, in general, on
  $R_k$.}
\end{center}
\end{figure}

ERG flows are based on the Wilsonian idea of integrating out momentum
modes within a path integral representation of quantum field theory.
In its modern form, the ERG flow for an effective action $\Gamma_k$
for bosonic fields $\phi$ is given by the simple one-loop expression
\cite{Bagnuls:2000ae}
\beq\label{ERG} 
\partial_t\Gamma_k[\phi] =
\frac{1}{2}\Tr\left( \frac{\de^2\Gamma_k}{\delta \phi\delta\phi} +
  R_k\right)^{-1} \partial_t R_k 
\eeq 
Here, $\Gamma^{(2)}_k$ denotes the second functional derivative of the
effective action, $t\equiv\ln k$ is the logarithmic scale parameter,
and $R_k(q^2)$ is an infrared (IR) regulator at the momentum scale
$k$. The regulator $R_k$ obeys a few restrictions, which have been
discussed at length in the literature \cite{Bagnuls:2000ae}. They
ensure that the flow equation is well-defined, thereby interpolating
between an initial action in the UV and the full quantum effective
action in the IR.  In order to solve \eq{ERG}, we have to specify an
initial effective action $\Gamma_\Lambda$ at some ultraviolet (UV)
scale $k=\Lambda$, and a regulator $R_k$. Clearly, the flow trajectory
of \Eq{ERG} in the space of all action functionals depends on the IR
regulator function $R_k$.  For the full flow, this is of no relevance.
Starting from an initial effective action $\Gamma_\Lambda$, the
integrated full flow approaches the full quantum effective action,
independently on the choice for $R_k$ along the flow. Schematically,
this is depicted in Fig.~\ref{fig:flows}a.

The situation changes once truncations have been made. Here,
``truncations'' mean that some vertex functions are neglected in the
Ansatz for the functional form of $\Gamma_k^{\rm trunc}$ entering
\eq{ERG}.  Schematically, this scenario is depicted in
Fig.~\ref{fig:flows}b.  Still, the flow trajectories in the space of
all action functionals depend on $R_k$.  However, it cannot be
guaranteed that the endpoint of the integrated flow is independent on
$R_k$. In general, it is not.  The origin of this spurious scheme
dependence is easily understood: while regulating the flow, the
regulator $R_k$ also modifies all vertex functions and their
interactions at $k\neq 0$.  Hence, the ``missing'' back-coupling of
neglected vertex functions is responsible for a spurious scheme
dependence. An immediate consequence of this observation is that
varying the regulator influences the physical content of a given
truncation.  Hence, the scheme dependence within a given truncation,
and convergence properties of ERG flows are entangled \cite{Litim:2000ci}.

\section{Optimisation and stability}\label{Optimisation}

Next we turn to the stability of the flow \eq{ERG}, and a simple
optimisation condition \cite{Litim:2000ci,Litim:2001up}. The two
ingredients of \eq{ERG} are the full regularised propagator
$(\Gamma^{(2)}_k+R_k)^{-1}$ -which contains the physical information
of the flow-, and the insertion $\partial_t R_k$.  Typically,
$\partial_t R_k$ is peaked around $q^2\approx k^2$, and decays
exponentially for large momenta. For small momenta, the flow \eq{ERG}
is regularised due to $R_k$ in the full propagator. The regulator
implies that the inverse propagator displays a gap,
\beq\label{gap}
\min_{q^2\ge 0} 
\left(\left.
\0{\delta^2\Gamma_k[\phi]}{\delta\phi\, \delta\phi}
\right|_{\phi=\phi_0}+
 R_k \right) > C\,k^2
\eeq
with $C>0$. The minimum is achieved for $q^2\approx k^2$. In general,
$C$ depends on $R_k$ and on $\phi_0$. The flow \eq{ERG} receives its
dominant contributions from the region in momentum space where
$\partial_t R_k$ is large and the inverse propagator is small. In
consequence, the flow is more stable against small changes in
$\Gamma_k$ the larger the full inverse propagator.  This observation
leads to a simple criterion to optimise the stability of flows. To
that end, let us consider a theory with a standard propagator and
$\Gamma^{(2)}(\phi)=q^2 + U_k''(\phi)$. This corresponds to the
leading order in a derivative expansion. Inserting this expression
into \eq{gap}, we require the gap to be maximal w.r.t.~the
regularisation scheme. Dropping irrelevant momentum-independent terms,
the optimisation condition becomes
\beq\label{opt}
\max_{\rm (RS)}\left[
\min_{q^2\ge 0} 
\left(q^2 +R_k(q^2)\right) 
\right]
\Rightarrow R_{\rm opt}
\eeq
for any fixed $k$. Eq.~(\ref{opt}) states that an optimised regulator
maximises the gap \eq{gap} w.r.t.~the regularisation scheme (RS)
\cite{Litim:2000ci}.  The condition is based only on properties of the
flow \eq{ERG}, and not on the specific theory under investigation. To
leading order in the derivative expansion, solutions to the condition
\eq{opt} are independent on the specific theory. In general, solutions
$R_{\rm opt}$ to \eq{opt} are not unique and depend on the class of
regulators chosen for the optimisation. Still, it is worthwhile
noticing that \eq{opt} is a rather mild condition: it fixes only one
out of countable infinitely many parameters describing a regulator
$R_k$ \cite{Litim:2000ci}.

We stress that the present considerations are based on the structure
of ERG flows of the form \eq{ERG}. Similar considerations can be
applied to other exact RG flows based upon momentum shell
integrations, like Wilsonian flows, the Polchinski RG,
Wegner-Houghton flows, Hamiltonian flows or generalised proper-time
flows \cite{Litim:2001ky}. In contrast, an implementation is less
transparent for RG flows based upon reparametrisation invariance.

As an example, we consider a scalar theory to leading order in the
derivative expansion, using a standard kinetic term. Higher order
corrections can be treated as well. Then, a simple solution to the
optimisation condition \eq{opt} is given by \cite{Litim:2001up}
\beq\label{Ropt}
R_{\rm opt}(q^2)=(k^2-q^2)\theta(k^2-q^2)\,.
\eeq
For momenta $q^2 > k^2$, it leads to
\beq\label{large}
\Gamma^{(2)}_k[\phi]+R_{\rm opt}(q^2) = q^2+U''_k(\phi)
\eeq
\Eq{large} states that the regularisation is absent for large momenta.
For $q^2< k^2$, we find
\beq\label{small}
\Gamma^{(2)}_k[\phi]+R_{\rm opt}(q^2)= k^2+U''_k(\phi)\,.
\eeq
In this region, the inverse propagator \eq{small} is ``flat'', {\it
  i.e.} independent of momenta.  Hence, all IR modes are treated
equally.  The regulator \eq{Ropt} has a number of interesting
properties \cite{Litim:2001up}. It leads to the fastest decoupling of
heavy modes, it disentangles the contribution of quantum and thermal
fluctuations along the flow, it leads to a factorisation of a
homogeneous wave function renormalisation, it leads to a smooth
approach to convexity for a theory in the phase with spontaneous
symmetry breaking, and it improves the convergence of the derivative
expansion \cite{Litim:2001dt}. The link to a minimum sensitivity
condition has been established as well \cite{Litim:2001fd}. Finally,
the choice \eq{Ropt} is also useful from a technical point of view,
because it leads to a simple analytic flow.

More generally, (most of) these properties hold as well for other
optimised flows, different from \eq{Ropt}, as long as \eq{small} holds
approximately in the momentum region where the flow receives its
dominant contributions.  We restricted the discussion to bosonic
fields. Extensions to fermions and gauge fields have been considered
as well \cite{Litim:2001up}.

\section{Stability and convergence}\label{Stability}

In the remaining part, we apply our reasoning for $O(N)$-symmetric
real scalar theories at the Wilson-Fisher fixed point in $d=3$
Euclidean dimensions.  The universality class is characterised by the
critical exponent $\nu_{\rm phys}$, given by the inverse of the
negative eigenvalue of the stability matrix at criticality, and
$\eta_{\rm phys}$, the anomalous dimension.  It is known from
experiment that $\eta_{\rm phys}$ is at most of the order of a few
percent.  Hence, it is believed that the derivative expansion is a
good approximation for a reliable computation of universal critical
exponents.  Within the derivative expansion, the physical critical
exponents at the scaling solution are computed as the series
\bea\label{nu-expansion}
\nu_{\rm phys} =&\nu_{0}({\rm RS})&  + \nu_{1}({\rm RS})
                                     + \nu_{2}({\rm RS})+\cdots
\eea
Here, the index corresponds to the order of the derivative expansion.
The anomalous dimension $\eta$ vanishes to leading order.
Notice that every single order in the expansion --- due to the
approximations employed --- depends on the regularisation scheme.  The
independence of physical observables on the regulator scheme (RS) can
only be guaranteed in the limit where {\it all} operators of the
effective action are retained during the flow.  In turn, the physical
value $\nu_{\rm phys}$ is independent of the precise form of the
infrared regulator. Hence, the infinite sum on the right-hand side
adds up in a way such that the physical values are scheme independent.
The convergence of the expansion \eq{nu-expansion} is best if a
regulator is found such that the main physical information is
contained in a few leading order terms.

\begin{figure}[t]
\begin{center}
\unitlength0.001\hsize
\begin{picture}(1000,900)
\put(230,470){\large $n_{\rm trunc}$}
\put(730,470){\large $n_{\rm trunc}$}
\put(300,850){$\rho = 0$}
\put(800,850){$\rho=\rho_0$}
\put(280,600){\fbox{\large $\nu_{\rm Ising}$}}
\includegraphics[width=\hsize]{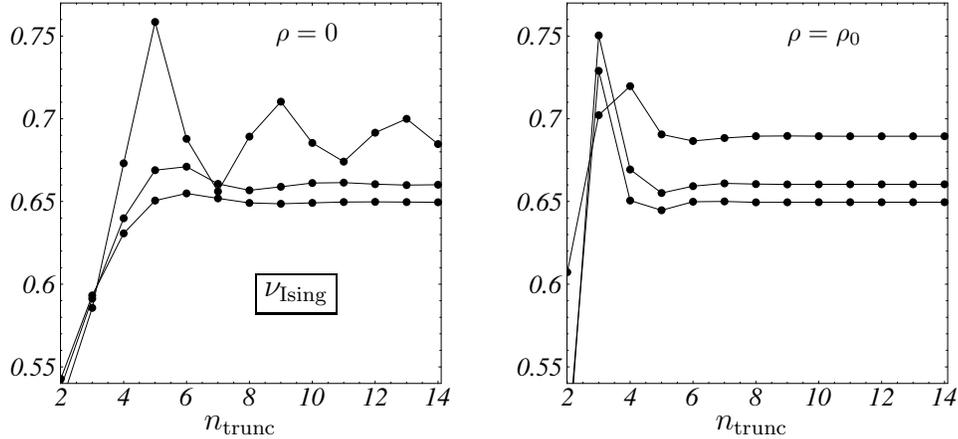}
\end{picture}
\vskip-6.5cm
\caption{\label{fig:convergence} The critical index $\nu$ for the Ising
  universality class. Results are given for an expansion about
  $\rho=0$ (left panel) and $\rho=\rho_0$ (right panel), and for the
  sharp cutoff (upper curves) the quartic regulator (middle curves)
  and the optimised regulator (lower curves). The stability is largely
  improved by replacing the non-optimised sharp cutoff by optimised
  ones.}
\end{center}
\end{figure}

In Fig.~\ref{fig:convergence}, we have computed $\nu$ for the Ising
universality class within a polynomial approximation for the scaling
potential, using a sharp cutoff $R_{\rm sharp}$ (upper curves), the
quartic regulator $R_{\rm quart}=k^4/q^2$ (middle curves), and $R_{\rm
  opt}$ (lower curves) \cite{Litim:2001dt,Litim:2002cf}.  Both $R_{\rm
  quart}$ and $R_{\rm opt}$ are optimised regulators [solutions to
\eq{opt}], while $R_{\rm sharp}$ is not. For the left panel, we have
expanded the scaling potential in polynomials of $\rho\equiv \phi^2/k$
around vanishing field up to order $n_{\rm trunc}$. For the right
panel, the expansion has been performed around the local minimum
$\rho=\rho_0$.  A few lessons can be learnt from
Fig.~\ref{fig:convergence}: 

First, it is seen that the convergence and stability of the sharp
cutoff flow is poor. The expansion depends strongly on the expansion
point. For an expansion about vanishing fields, it does not even
converge beyond a certain accuracy \cite{Morris:1994ki,Aoki:1998um}.
In contrast, the polynomial approximation converges rapidly for both
$R_{\rm quart}$ and $R_{\rm opt}$. Also, the convergence depends only
weakly on the expansion point. This picture holds true for any $N$.
These findings confirm that optimised flows are more stable. In this
light, the non-convergence of the sharp cut-off flow within an
expansion about vanishing field is considered as a deficiency of the
sharp cut-off regularisation, and not of the expansion.

Second, we notice that the numerical values for the critical exponent
$\nu$ depends on the regulator. In particular, the values obtained
from optimised flows are closer to the physical value. This holds true
for all $N\ge 0$ \cite{Litim:2001dt,Litim:2002cf}. Based on an
investigation of a large class of regulator functions, it has even
been argued that the value $\nu_{\rm opt}$ as obtained from \eq{Ropt}
corresponds to a minimum \cite{Litim:2002cf},
\beq\label{bound}
\nu_{{\rm large}-N}\ge\nu_{{}_{\rm ERG}}\ge\nu_{\rm opt}>\nu_{\rm phys}\,.
\eeq
Here, the upper bound denotes the large-$N$ limit, for all $N$. The
result \eq{bound} shows that the regulator \eq{Ropt} correponds to a
solution of a minimum sensitivity condition \cite{Litim:2001fd}.

Third, it is interesting to note that $\nu_{\rm opt}$ agrees to all published
digits with the results obtained from the Polchinski RG
\cite{Ball:1995ji,Comellas:1997tf}. This is remarkable insofar as the
two flows, ERG and Polchinski RG, are related by a Legendre transform
and appropriate field rescalings.  Hence, their derivative expansions
are not equivalent. Also, to leading order, the result from a
Polchinski flow is scheme independent, in marked contrast to what has
been found within the ERG.  The agreement between the Polchinski RG
result and the ERG result based on \eq{Ropt} suggests that the
optimisation has removed a redundant scheme dependence from the ERG
flow.

\begin{figure}
\begin{center}
{}\vskip-1.cm
\unitlength0.001\hsize
\begin{picture}(600,580)
\put(200,380){ \fbox{{ $\displaystyle 
\log_{10}\left|\0{\nu_{\rm opt}-\nu_{\rm trunc}}{\nu_{\rm opt}}\right|$}}}
\put(230,-10){\large $n_{\rm trunc}$}
\includegraphics[width=.5\hsize]{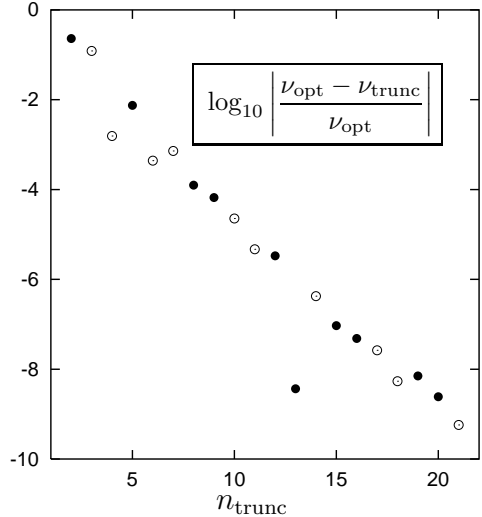}
\end{picture}
\vskip.3cm
\caption{\label{fig:pVergleich3} Ising universality class.  Convergence of
    $\nu_{\rm trunc}$ towards $\nu_{\rm opt}$ with
    increasing truncation. Points where $\nu_{\rm trunc}$ is larger
    (smaller) than $\nu_{\rm opt}$ are denoted by o ($\bullet$).}
\end{center}
\end{figure}

Next, we emphasize that the numerical convergence of $\nu_{\rm trunc}$
from the optimised flow towards $\nu_{\rm opt}$ is very fast
(Fig.~\ref{fig:pVergleich3}): typically, increasing $n_{\rm trunc}$ by
$2-2.5$ increases the numerical accuracy by one decimal point. Given
that the accuracy of $\nu$ cannot be better than a few percent
(contributions $\sim\eta$ are suppressed to leading order in the
derivative expansion), it suffices to retain $\nu_{\rm trunc}=4 (6)$
independent couplings in the Ansatz for the effective potential, in
order to achieve an accuracy for $\nu_{\rm trunc}$ below $1\%$
($0.1\%$). This efficiency is remarkable.

\begin{figure}[t]
\begin{center}
\unitlength0.001\hsize
\begin{picture}(700,600)
\put(330,500){
\begin{tabular}{ll}
Sharp&${}^{\multiput(0,0)(20,0){4}{\line(10,0){10}}} $\\[-.7ex] 
Quart&${}^{\multiput(0,0)(20,0){3}{\put(0,0){\line(10,0){10}}
\put(14,0){\line(2,0){2}}}\put(60,0){\line(10,0){10}}}${}\\[-.7ex] 
Opt&$  {}^{\put(0,0){\line(70,0){70}}}${}
\end{tabular}}
\put(515,120){$\infty$}
\put(300,80){ {\large $N$}}
\put(370,400){ \fbox{{\large $\displaystyle \0{\nu_{{}_{\rm \tiny ERG}}}{\nu_{\rm opt}}-1$}}}
\hskip.04\hsize
\includegraphics[width=.5\hsize]{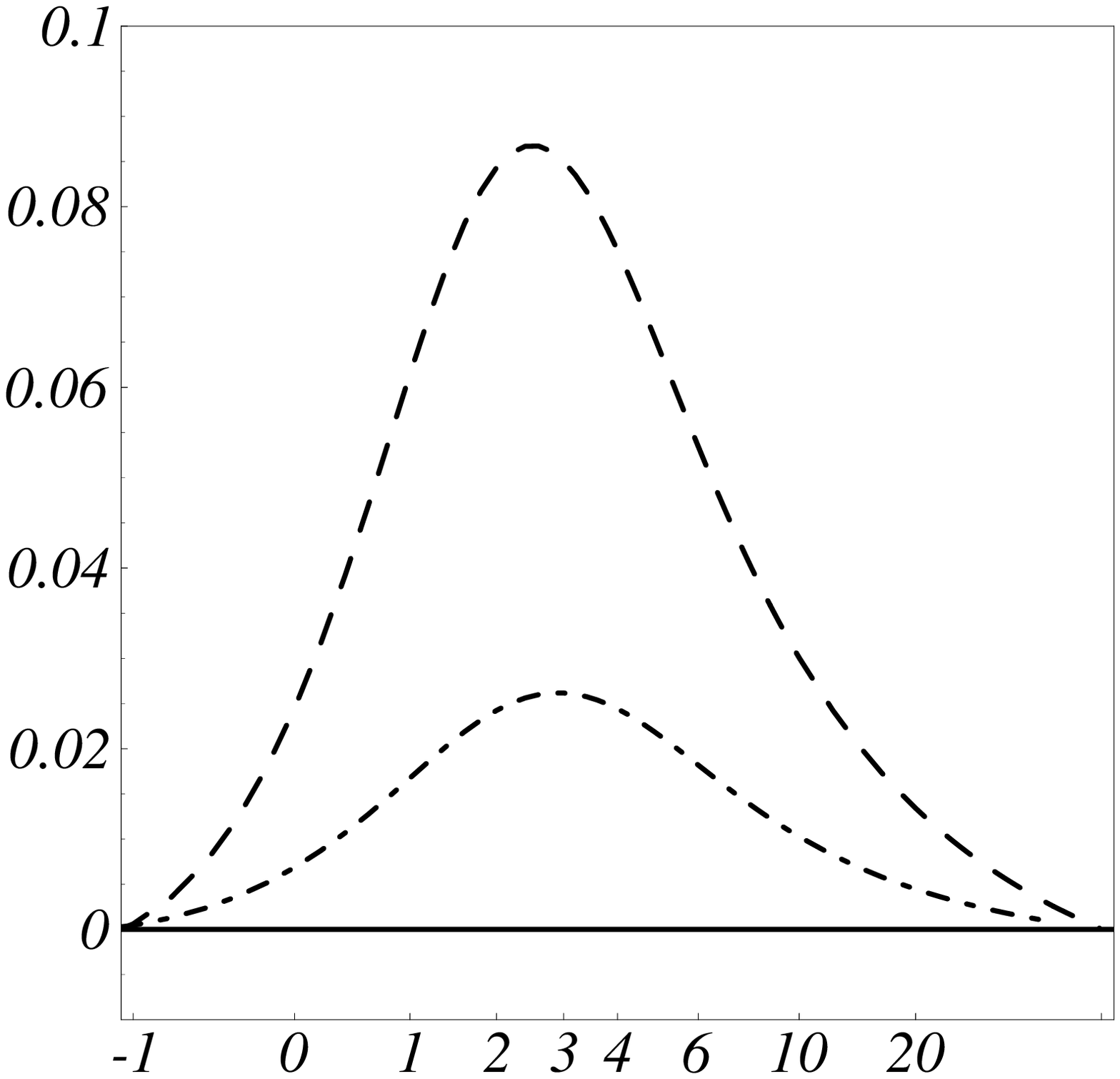}
\end{picture}
\vskip-1.3cm
\caption{\label{fig:pVergleich2} The relative improvement 
of $\nu_{{}_{\rm \tiny ERG}}$ in comparison with $\nu_{\rm opt}$
  (various regulators). }
\end{center}
\end{figure}

Finally, we discuss in Fig.~\ref{fig:pVergleich2} the relative
improvement due to an optimised regulator for all $N$
\cite{Litim:2001dt}. For comparison, we again took the sharp and the
quartic cutoff. It is interesting to note that both the large-$N$
limit and the case $N=-2$ lead to universal leading-order results for
$\nu$ \cite{BT:1973,Litim:1995ex}. For intermediate values, the
results deviate significantly from the best one, up to nearly $9-10\%$
for the sharp cutoff, and $2-3\%$ for the quartic one. An improvement
by up to $10\%$ is very important, given that the physical value lies
a few percent below the values found for $\nu_{\rm opt}$ (for all $N$
of physical relevance). Hence, for flows based on the sharp cutoff or
similar regulators, one expects that a higher order in the derivative
expansion is required to achieve the same accuracy in comparison to
optimised flows. For a more detailed discussion of the link between
the convergence of the derivative expansion, and the optimisation, we
refer to the discussion in \cite{Litim:2001dt}.

\section{Conclusions}\label{Conclusions}

Within the framework of the ERG, we have studied the link between
stability and convergence properties of ERG flows, and their
dependence on the regularisation. This understanding is a prerequisite
for reliable applications of the formalism. In this context, the
exactness of the flow \eq{ERG} plays an important role. These
considerations have lead to a simple optimisation condition for ERG
flows. When applied to scalar theories at criticality in $3d$, we have
shown explicitly that the optimisation leads to improved results
already to leading order in the derivative expansion. This
understanding of the spurious scheme dependence has reduced the
ambiguity in $\nu$ to a small range about $\nu_{\rm opt}$. 

Some of our results are based on the particularly simple choice
\eq{Ropt} for the regulator. However, many more optimised regulators
are available, and other choices may even be more appropriate
depending on the order of the truncation, or on the physical problem
under investigation. This can be seen already from the present
results. To leading order in the derivative expansion, and as a
function of the regularisation, the critical index $\nu$ is very flat
\cite{Litim:2002cf}. Higher order corrections $\sim\eta$ are
subleading. However, it is expected that the (nearly) flat region for
$\nu$ is resolved to higher order in the derivative expansion, once
$\eta\neq 0$.

Based on the understanding achieved so far, we expect that optimised
flows should be useful for applications to higher in the derivative
expansion, or for applications to quantum gravity
\cite{QuantumGravity1}, to more complex scalar theories
\cite{Tissier}, to fermionic ones \cite{Gies:2002kd}, or to
non-Abelian gauge theories \cite{QCD1,Gies:2002af,Pawlowski}. In all
these cases, an implementation of the ERG is technically much more
demanding, and approximations are often bound to lower orders as
compared to (simpler) scalar theories. Therefore, it may be most
helpful to apply optimised flows and to achieve improved results
already to lower orders in the truncation.

\section*{Acknowledgements}
It is a pleasure to thank the organisers for the stimulating
conference. This work has been supported by the DFG, and by a
Marie-Curie fellowship under EC contract no.~HPMF-CT-1999-00404.


\end{document}